\documentclass[%
notitlepage,
twocolumn,
amsmath,amssymb,
]{revtex4-1}

\usepackage{graphicx}
\usepackage{stackengine}
\usepackage{dcolumn}
\usepackage{bm}
\usepackage{xcolor}
\usepackage{mathtools}
\usepackage{bbold}
\usepackage{subfigure}
\usepackage{mathtools} 

\usepackage[autostyle=true,german=quotes]{csquotes}

\newcommand{\dg}{\dagger}

\newcommand{\ket}[1]{| #1 \rangle}

\begin{document}
	
	\preprint{APS/123-QED}
	
	\title{Non-equilibrium non-Markovian steady-states in open quantum many-body systems:
		Persistent oscillations in Heisenberg quantum spin chains}
	
	\author{Regina Finsterh\"olzl}
	\email{regina.finsterhoelzl@tu-berlin.de}
	\author{Manuel Katzer}
	\author{Alexander Carmele}
	\affiliation{Technische Universit\"at Berlin, Institut f\"ur Theoretische Physik, Nichtlineare Optik und Quantenelektronik, Hardenbergstra{\ss}e 36, 10623 Berlin, Germany}
	
	\date{\today}
	
	\begin{abstract}
		We investigate the effect of a non-Markovian, structured reservoir on an open Heisenberg spin chain. 
		We establish a coherent self-feedback mechanism as the reservoir couples frequency-dependent to the spin chain.
		Thus, loss and driving take place due to the interaction of the spin chain with its own past. 
		This new paradigm of non-Markovian imposed boundary-driving allows to discuss a new kind of non-equilibrium steady-state. We show that for certain parameters even in the long-time limit persistent oscillations occur within the chain. 
		Moreover, we demonstrate that the conditions for these oscillations and excitation trapping depend on the characteristics of the chain, thus making it possible to characterize a chain by detection of its emitted signal under influence of self-feedback.
		%
		%
	\end{abstract}
	
	\maketitle
	
	\section{Introduction}
	Quantum spin chains are a paradigm to study quantum many-body physics out-of-equilibrium \cite{Rotter2015,Eisert2015,Breuer2002,Levi2016,Buyskikh2019,Owen2018} and exhibit a rich variety of dynamical properties such as phase transitions \cite{Droenner2017,Znidaric2016,Heyl2013,Huber2019,Huber2019active,Pizzi2020}, quantum transport properties \cite{Bertini2020,Hauke2013, Trautmann2018,Prosen2011,Ljubotina2017,Lange2018,Znidaric2011,Znidaric2011a,Katzer2020} and entanglement structure \cite{Wang2007,Wu2011}.
	Among quantum spin chains, the Heisenberg spin-$1/2$ chain \cite{Heisenberg1985} is particularly important as it is analytically solvable \cite{Bethe1931,Dupont2020} and forms the backbone to explain experiments in the domain of strongly-correlated many-body physics \cite{Hild2014,Tang2018,Langen2015,Kinoshita2006,Maier2019}.
	Part of this research in open quantum systems focuses on a spin chain which is coupled to magnetic reservoirs at both ends \cite{Prosen2015,Prosen2009,Karevski2013,Cai2013,Xu2018,Mendoza-Arenas2019,Popkov2019}.
	Based on a full Markovian approximation with respect to the system-reservoir interaction, the chain is incoherently driven into a non-equilibrium steady state and the influence of the driving strength via the external reservoir, of an externally-induced disorder parameter \cite{Droenner2017,Katzer2020,Znidaric2016} or the strength of the anisotropy \cite{Karrasch2014,Ilievski2018,Medenjak2017} are discussed.

\begin{figure}[b!]
		\includegraphics[width=0.8\linewidth]{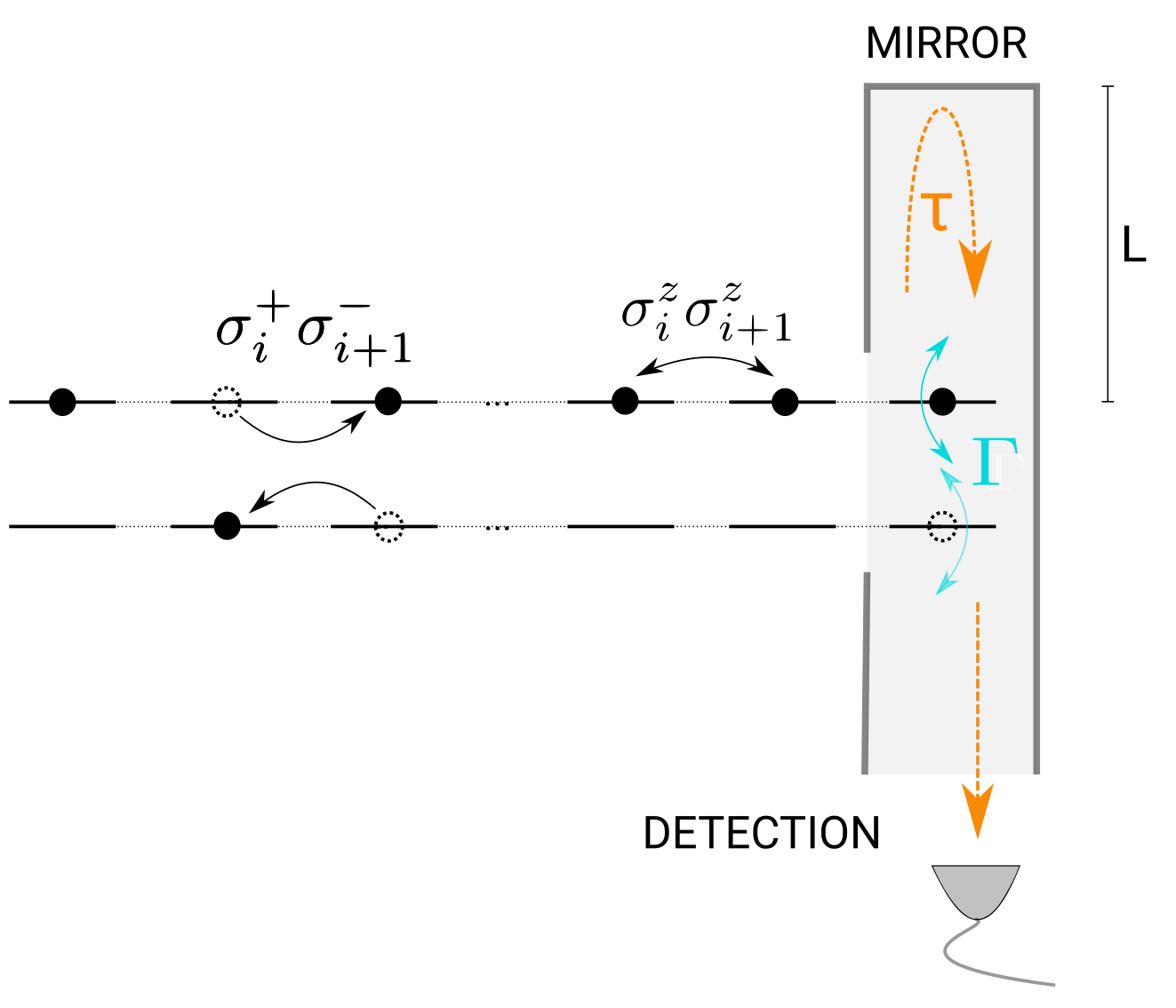}
		\caption{Sketch of a Heisenberg spin chain modeled as coupled two level systems with the coupling strength $J$. The last site couples with the rate $\Gamma$ to a reservoir consisting of a semi-infinite waveguide of length $L$ at the closed end, which feeds back part of the excitation after a delay time $\tau=2L/c$. At the open end of the waveguide, a detector records the emitted signal for a period of duration $T$.}
		\label{fig1}
	\end{figure}

	Complementing this Markovian, Lindblad-based approach to describe a boundary-driven quantum spin chain, we investigate in the present study the effect of a non-Markovian, structured reservoir on an open Heisenberg chain \cite{Altafini2007,Morigi2015,Roos2019}.
	The structured reservoir couples frequency-dependent to the spin chain and therefore introduces a memory.
	Here, we choose a $\delta-$like memory kernel to establish a coherent self-feedback mechanism \cite{lloyd2000coherent,Wiseman1994}, i.e. the spin chain interacts partially with its own past and the boundary-driven setup is changed from a spatial to temporal-driving scheme: Loss and driving take place at the same site but include two different points in time separated by the roundtrip-time $\tau=2L/c$, cf. Fig.~\ref{fig1}. 
	This new paradigm of non-Markovian imposed boundary-driving allows to discuss a new kind of non-equilibrium steady-state: The dissipative coupling to the structured reservoir leads for certain parameters to stabilized and non-decaying, i.e. persistent oscillations within the chain. 
	Since for these parameters the excitation in the chain remains constant and the amplitudes exhibit a regular oscillation pattern, this feature is related to Rabi oscillations which are intrinsically coherent and time-reversible.

	The enabling factor in our scheme is the non-Markovian system-reservoir coupling based on coherent feedback known from and predominantly studied in atom-molecular-optics and cavity-QED \cite{dorner2002,Cook86feedback,milonni,cook1987quantum,Cook2018,Nemet2016,Faulstich2018,Pichler2016,crowder2020quantum,Barkemeyer2020}. 
	Its coherent and non-Markovian nature introduces quantum interferences into the dynamics of these systems and allows for interesting two-photon processes \cite{Droenner2019,Pichler2016}, enhanced entanglement and non-classical photon statistics \cite{lu2017intensified}, dimerization \cite{guimond2017delayed,guimond2016chiral} and a stabilization of quantum coherence due to interference effects between incoming and outgoing probability waves \cite{Carmele2013}.
	Together with the formation of dark states and subsequently emerging population trapping \cite{Nemet2019,carmele2020pronounced}, Rabi oscillations in the single-excitation regime has been predicted \cite{Carmele2013}.
	These cavity-induced Rabi oscillations emerge if the roundtrip-time $\tau$ is a multiple of the inverse of the cavity-emitter coupling $g/(2\pi)$. 
	They are up-to-now limited to the single-excitation and single-emitter regime.

	Here, we show that these limitations can be lifted and the phenomenon of feedback-induced stabilization of Rabi oscillations is of general character and applies also to strongly-correlated many-body systems such as the Heisenberg chain.  
	In the following, we show that for certain parameters, it is possible to stabilize highly symmetric states within the chain depending on the feedback time. We propose thereby a way to control the state of the chain non-invasively and show that for the isotropic Heisenberg spin chain with nearest neighbor interaction the number of possible trapping conditions is equal to the number of sites in the chain.
	This allows for a partial characterization of the spin chain by its emitted, detector-integrated signal and extends the feedback-phenomenon of stabilized Rabi oscillations to the realm of strongly-correlated open quantum many-body systems. 

	This paper is organized as follows: First, in Sec.~\ref{sec:model}, we present the system of the spin chain and the numerical implementation of the feedback interaction. We realize the coherent self-feedback by placing the end of the spin chain in a semi-infinite waveguide which induces a frequency-dependent partial interaction of the spin chain with its own past after the roundtrip-time $\tau=2L/c$.
	In this section, we also explain the tensor network method we use for our numerical simulations: The quantum stochastical Schr\"odinger equation serves as the basis for a efficient description of the time evolution with matrix product states  \cite{Pichler2016,Droenner2019,carmele2020pronounced}.
	Next, in Sec.~\ref{sec:without_feedback}, we discuss the system behavior without feedback and find that in our setup, no population trapping can occur and dark states cannot be populated.
	This is in contrast to the feedback case, investigated in Sec.~\ref{sec:with_feedback}, where we find pronounced population and persistent oscillations. We study the conditions for population trapping and show that strikingly, despite the complex many-body interactions within the chain, the number of trapping conditions is equal to the number of sites in the chain. Investigating the Rabi oscillations we find that the amplitude is highest for a single excitation in the chain. We conclude in Sec.~\ref{sec:conclusion} and give a short outlook of possible applications of our scheme.
	%
	
	\section{System}\label{sec:model}
	Our model consists of a Heisenberg spin chain whose last site is coupled to a non-Markovian structured reservoir,cf. Fig.~\ref{fig1}. 
	This reservoir is created via a semi-infinite waveguide \cite{Hughes2007,Fang2015,calajo,dorner2002} where the closed end is modeled by a mirror in distance $L$ to the spin chain.
	The reservoir is assumed to be initially in the vacuum state. 
	Part of the excitation emitted from the chain will then be reflected by the mirror and interacts with the system a second time after the delay time $\tau$. 
	While this model is well investigated for a single few-level emitter \cite{Barkemeyer2020,Carmele2013,guimond2016chiral,Pichler2016,crowder2020quantum,carmele2020pronounced,Nemet2016,Nemet2019}, we extend the investigation here to a many-body system. 

	The corresponding Hamiltonian of the combined system-reservoir dynamics reads (with $\hbar \equiv 1$):
	\begin{align}
	H = 
	&\sum_{i=1}^N \omega_0 \sigma_i^+\sigma_i^- + \int d\omega\, \omega b^\dagger (\omega) b (\omega)\nonumber\\
	&+\sum_{i=1}^{N-1} J \big(\sigma^x_i\sigma^x_{i+1}+\sigma^y_i\sigma^y_{i+1}
	+\sigma^z_i\sigma^z_{i+1}\big)\nonumber\\
	&+\int d\omega \left( G_{fb} (\omega) b^\dagger_N(\omega) \sigma_N^- + \textup{h.c.} \right)
	\label{equ:model}
	\end{align}
	The first term models the free evolution of $N$ single spin systems, where $\omega_0$ governs the free evolution of each single site, $\sigma_i^+=\sigma_i^x + i \sigma_i^y$ and $\sigma_i^-=\sigma_i^x - i \sigma_i^y$ create/annihilate a fermionic excitation in the $i$th two-level system which is equivalent to a flip of the spin on site $i$ \cite{Wang2020,carmele2020pronounced,Pichler2015spin,Ramos2014,Ramos2016}.
	The second term represents the free evolution of the bosonic mode continuum to which the last site is coupled. Here, $b^{(\dagger)}_N(\omega)$ creates/annihilates a bosonic excitation of energy $\omega$ in interaction with the $N$th site of the spin chain.
	The third term models the isotropic Heisenberg spin chain with nearest neighbor interaction, a chain of $N$ single sites and with a three-dimensional nearest neighbor interaction in $x$, $y$ and $z$ direction, where $\sigma^k$, $k \in {x,y,z}$ represent the Pauli matrices interacting with strength $J$. 
	The last term represents the interaction of the $N$th site of the chain with the bosonic reservoir and offers a unitary description of decay and feedback effects by interaction with the reservoir. 
	The system-reservoir coupling $G_{fb}(\omega)$ is sinusoidal frequency dependent in order to model a semi-infinite waveguide \cite{Trautmann2016,Faulstich2018,Cook2018,Cook86feedback,Tufarelli2013,Tufarelli2014}:
	\begin{equation}
	G_{fb} (\omega) = g_0 \sin{\left(\frac{\omega L}{c_0} \right)} = i \sqrt{\frac{\Gamma}{2\pi}} \left( e^{-i\omega \tau/2}- e^{i\omega \tau/2}\right)
	\end{equation}
	where $L$ is the length of the closed side of the waveguide, $c_0$ the phase velocity in the waveguide, $\tau=2L/c_0$ the delay time and $g_0=\sqrt{\Gamma/2\pi}$ the coupling constant with the coupling rate $\Gamma$.
	Due to this frequency-dependent coupling to the reservoir, the dynamics is simulated in the time-discrete quantum stochastic Schr\"odinger equation (QSSE) approach \cite{Pichler2016}. 
	In order to achieve this, tensor network methods are employed by describing the state of the system and of the reservoir numerically as a matrix product state (MPS). 
	Instead of tracing out the reservoir's degrees of freedom, we
	remain in the Schr\"odinger picture and use a time discrete basis which includes the interaction with the reservoir at one time step with a stochastical, time-stroboscopic description. 
	The time-ordered evolution operator 
	\begin{equation}
	U(t)=\hat{T}\exp\left(-i\int_{t_0}^{t}H'' (t') dt'\right).
	\end{equation}
	is expressed in a time-discrete basis with commutating operators for different time steps, cf.~App.~\ref{app:qsse} for details.
	The operators act on the reservoir at the time $t_k = k\Delta t$ with equidistant time steps $\Delta t = t_{k+1}-t_k$. The wavevector reads:
	\begin{align}
	\ket{\psi(t_k)}=\nonumber &\sum_{\stackrel{n_1 \dots n_N}{= 0,1}} c_{n_1 \dots n_N} \ket{n_1\dots n_N} \\
	&\otimes 
	\sum_{k_1 \dots k_{N_T}} c_{k_1 \dots k_{N_T}} \ket{k_1\dots k_{N_T}}
	\end{align}
	with the expanded coefficients written with tensors $A$, cf.~App.~\ref{app:mps} for details.
	This leads, together with unitary transformations, cf.~App.~\ref{app:rotatingframe}, to the following discretized time evolution operator:
	\begin{align}
	U&(t_{k+1},t_k)= \nonumber\\
	&=\exp\Bigg[\sum_{i=1}^{N-1} J\Big(\sigma^x_i\sigma^x_{i+1}+\sigma^y_i\sigma^y_{i+1}
	+\sigma^z_i\sigma^z_{i+1}\Big)\nonumber\\
	&+\sqrt{\Gamma} \Big(\Delta B_N(t_{k})-\Delta B_N(t_{k-l}) e^{i\phi} \Big)\sigma_N^+ \nonumber\\
	&-\sqrt{\Gamma} \left(\Delta B^\dagger_N(t_{k}) - \Delta B^\dagger_N(t_{k-l}) e^{-i\phi} \right)\sigma_N^-\Bigg]
	\label{eq:u}
	\end{align}
	for $k\in[0,N_T-1]$ as integer of the time steps.
	Here, $t_k$ denotes the $k$th time step, while $t_{k-l}=(k-l)\Delta t$ denotes the state of the reservoir at the time $t_k-\tau$ and $\phi=\omega_0\tau$ denotes the feedback phase.
	For details, please refer to App.~\ref{app:rotatingframe}-App.~\ref{app:mps}.

	\begin{figure}[b!] 
		\centering
		\includegraphics[width=0.5\textwidth]{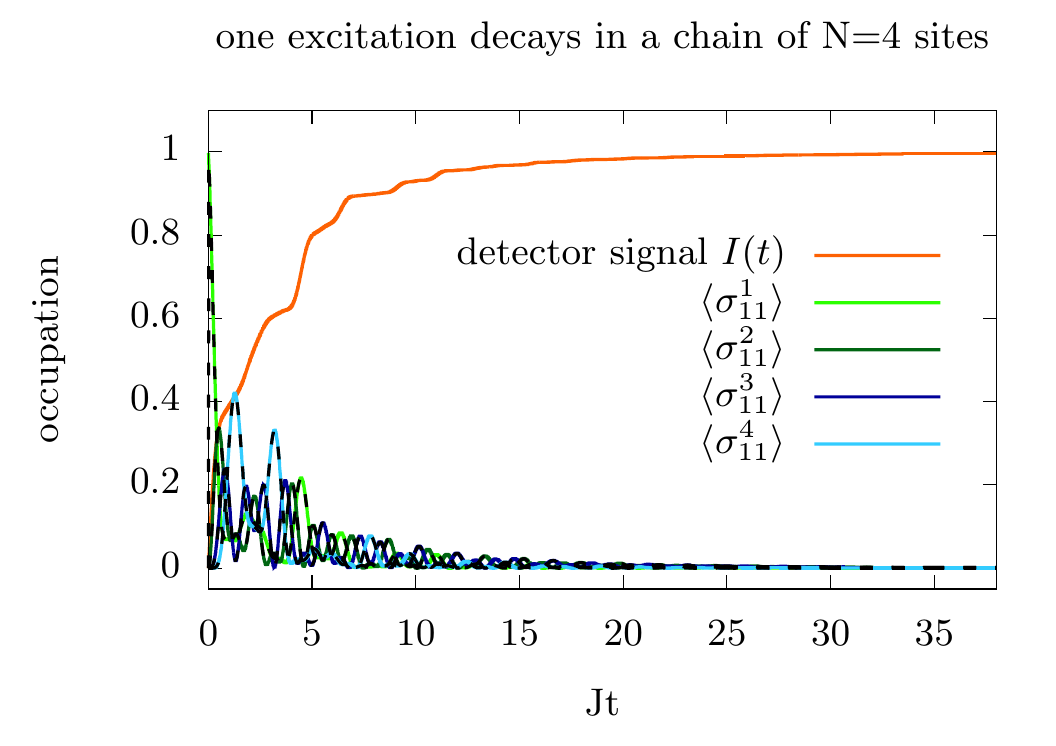}
		\caption{Time-dependent occupation densities in a Heisenberg chain of $N=4$ sites without feedback. Clearly, the initial state quickly dissipates into the environment and no excitations remain within the chain. Note that each curve is plotted twice demonstrating our benchmark. The orange line depicts the time-dependent detector signal which reaches its normalized maximum value after the convergence time $T_c$, thus $I(t=T_c)=1$. Parameters for this plot are $\Gamma=0.24$ and $J=0.1$.}
		\label{fig2}
	\end{figure}
	Due to our choice of a time-discrete basis, cf.~Eq.~\eqref{equ:basis}, the corresponding Hilbert space scales with the integration time and thus becomes very large. 
	In order to compute the time evolution we make use of the tensor network method based on matrix product states (MPS) called tMPS \cite{Schollwock2005,Schollwock2011,Vidal2003,Orus2008,White2004,White1993,Verstraete2004,Paeckel2019}.
	This method allows for an efficient truncation of the Hilbert space and has already successfully been applied on the time evolution of open spin systems \cite{Prosen2015,Prosen2009,Karevski2013,Cai2013,Xu2018,Mendoza-Arenas2019,Popkov2019,Mascarenhas2015} as well as of self-feedback problems for few-level systems, for instance for the simulation of quantum dots or cavity-embedded two level systems \cite{carmele2020pronounced,Nemet2019,Droenner2019}.

	Central to this method is the expansion of the state vector coefficient into a matrix product state, cf.~App.~\ref{app:mps}. While for low dimensional few-level systems, the state of the system and the reservoir may be written into one single MPS, we model the many-body system using a two dimensional MPS.
	In addition to the non-Markovian reservoir, our model also involves the spin chain as a quantum many-body system with spacial interaction, cf.~App.~\ref{app:algorithm}. This algorithm enables us to efficiently simulate a quantum many-body system under the influence of coherent self-feedback, i.e. a non-Markovian system-reservoir coupling.
	%
	
	\section{The dissipative Heisenberg chain without feedback} \label{sec:without_feedback}
	%
	First, we describe the Heisenberg chain dynamics without feedback. 
	In this case, only the boundary spin of the chain is subject to dissipation, i.e. it is coupled to a vacuum reservoir with vacuum input for every time step. 
	This is completely equivalent to a Markovian description with the Lindblad formalism.
	To benchmark the implementation, we have calculated the dynamics of the dissipative Heisenberg chain for the case of a vanishing frequency-dependence $G_{fb}(\omega)=2g_0$. 
	Therefore, the QSSE evolution models the Lindblad master equation of the form ($\hbar=1$):
	\begin{align}\label{eq:meq}
	\frac{\text{d}}{\text{dt}}
	\rho(t)
	&=-i\left[H_\text{chain},\rho(t)\right]
	+\Gamma \mathcal{D}[\sigma^-_N]\rho(t) \\
	H_\text{chain}
	&=
	\sum_{i=1}^N \omega_0 \sigma_i^+\sigma_i^- \notag \\ 
	&+\sum_{i=1}^{N-1}J\Big(\sigma^x_i\sigma^x_{i+1}+\sigma^y_i\sigma^y_{i+1}
	+\sigma^z_i\sigma^z_{i+1}\Big)
	\end{align}
	with the Lindblad superoperator $\mathcal{D}[J]\rho=J\rho J^\dg-J^\dg J\rho-\rho J^\dg J$.
	Note that with no feedback applied, our time-bin setting exactly reproduces the dynamics of a Lindblad decay, cf. Fig.~\eqref{fig2}.
	In this setting, excitation trapping is not possible for any initial state or parameter set, which means the excitation stored within the chain is inevitably lost to the reservoir modes.
	In our setup, we place a detector at the open end of the waveguide and record the time-dependent excitations which leave the feedback loop between the emitter at the end of the quantum chain and the mirror until we reach a finite time $T$. We time-integrated this excitation to form our detector signal $I(t)=\sum_{t_k=1}^{N_T} \langle \Delta B^\dagger(t_k) \Delta B(t_k) \rangle$.
	This time-integrated signal serves as our figure of merit. In the case without feedback, it will always reach unity if integrated long enough.
	In Fig.~\ref{fig2}, the time evolution of all sites is depicted exemplary for a spin chain of four sites (blue and green lines). The system has been initialized in the $\ket{\downarrow\downarrow\downarrow\downarrow \uparrow}\otimes\ket{\text{vac}}$ state. 
	Additionally, the time dependent detector signal is plotted (orange line), which integrates the dissipated signal during the integration time. 
	Clearly, all sites decay completely into ground state, and the signal at the detector $I(t)$ reaches its normalized maximum value after the convergence time $T_c$, thus $I(t=T_c)=1$.
	In the given setup, no population trapping or non-trivial steady-state can occur.
	Also, dark states are not populated as only a single-site couples dissipatively to the reservoir.
	This picture would change completely if more site were coupled to the reservoir \cite{carmele2020pronounced}.
	Figure~\ref{fig2} furthermore serves as a benchmark using the full solution for $\ket{\psi(t)}$ with the Lindblad master equation (black dotted lines). We note that we also benchmarked the feedback algorithm for the uncoupled last site using an analytical solution for a single two-level system \cite{Kabuss2015,dorner2002,Nemet2019}.
	%
	
	\section{The Heisenberg chain under feedback} \label{sec:with_feedback}
	\subsubsection{Population trapping} \label{subsec:population_trapping}
	Contrary to the Markovian case, we observe population trapping when subjecting the chain to coherent self-feedback. 
	This means that the initial excitation within the chain dissipates partially into the reservoir until this process is stopped by the interaction with the feedback signal and modifies the dissipative coupling due to quantum interferences. 
	As a consequence, after a parameter-dependent time $T_c$, the system-reservoir interaction reaches a steady-state and dynamically traps the remaining excitation within the chain.
	From this time on, the signal at the detector ceases and longer integration times have no impact on the amount of detected excitation.
	The conditions for population trapping depend on two parameters: The delay time $\tau$ and the feedback phase $\phi$. 
	Importantly, the two parameters are not independent in this setup, as it holds that $\phi=\omega_0 \tau$.
	However, a microwave modulation of hyperfine-level may disentangle the feedback phase $\phi$ from the feedback time $\tau$ \cite{Barkemeyer2020}.
	In the following, we assume that the initial state of the chain is all spins are in their ground state but the spin coupled to the reservoir is in excited state. As the reservoir is in a vacuum state initially, we are in the single-excitation regime.
	However, our study and results are not limited to the 
	single-excitation regime, but also hold for more excitations, as we will discuss further below.

\begin{figure}[t!] 
		\centering
		\includegraphics[width=0.5\textwidth]{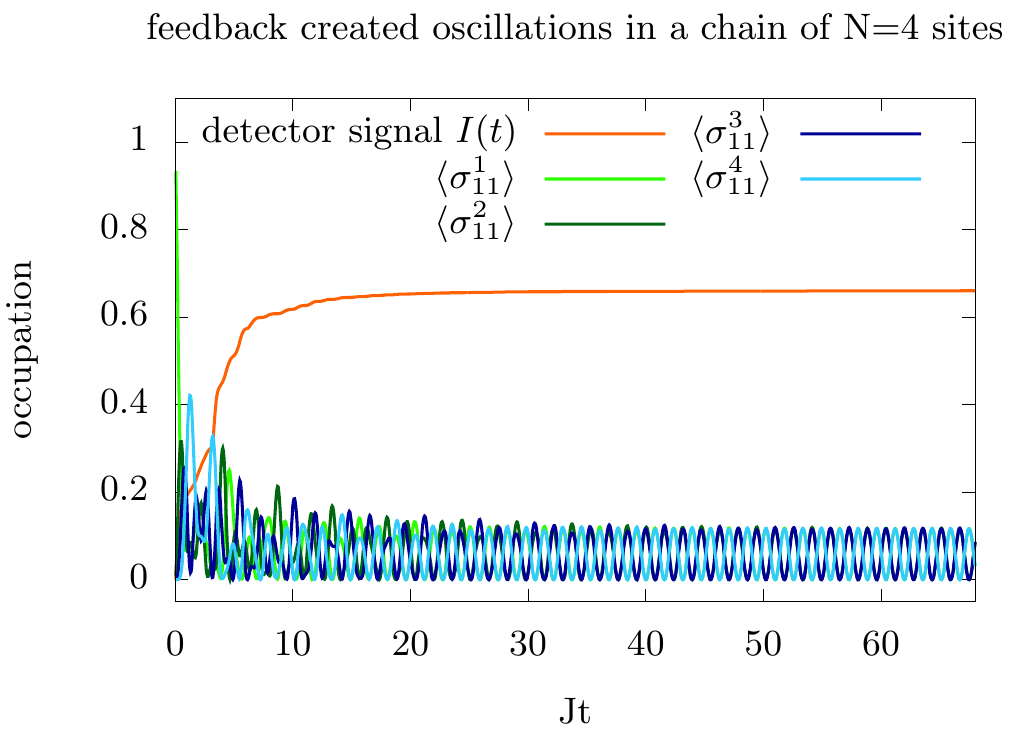}
		\caption{Time-dependent occupation densities $\langle \sigma_{11}^n (t) \rangle$ in a Heisenberg chain of $4$ sites. Clearly, feedback creates stable Rabi oscillations within the chain where site $2$ and $3$ as well as $1$ and $4$ are completely coherent and in phase. Consequently, part of the excitation remains trapped in the chain, clearly visible as the detector signal remains well below $I(T_c)=1$. As is explained below, these oscillations appear at intersection points of stability lines in the $\phi$-$\tau$ plane, where two trapping conditions are fulfilled at the same time. Parameters for this plot are $\Gamma=0.24$, $J=0.1$.}
		\label{fig3}
	\end{figure}

	In Fig.~\ref{fig3}, we plot the dynamics of the occupation densities in the Heisenberg chain of $N=4$ sites (blue and green lines) and the detected excitation leaving the waveguide (orange line). We show the transient regime as well as the long time limit.
	After a transient regime during which the densities within the chain oscillate irregularly and the detector signal steadily increases, the detector signal saturates and the densities within chain exhibit a very regular oscillation pattern.
	These oscillations are a special case of population trapping.
	Part of the initial excitation remains trapped in the chain and is swapped throughout the chain without any further losses. 
	Consequently, the detector signal cannot reach its maximum value. 

	This very unusual steady-state, in which the excitation within the feedback loop and within the chain are lossless swapped, and no excitation leaves the chain although the site couples dissipatively to a reservoir, is highly parameter dependent, as we will explain in the following.
	Namely, these oscillations appear at intersection points of stability lines in the $\phi$-$\tau$ plane, where two trapping conditions are fulfilled at the same time. 
	Also, we will show that strikingly, despite the complex many-body dynamics in the Heisenberg chain, the number of trapping conditions is equal to the number of sites in the chain.
	\begin{figure}[h!] 
		\centering
		\includegraphics[width=0.5\textwidth]{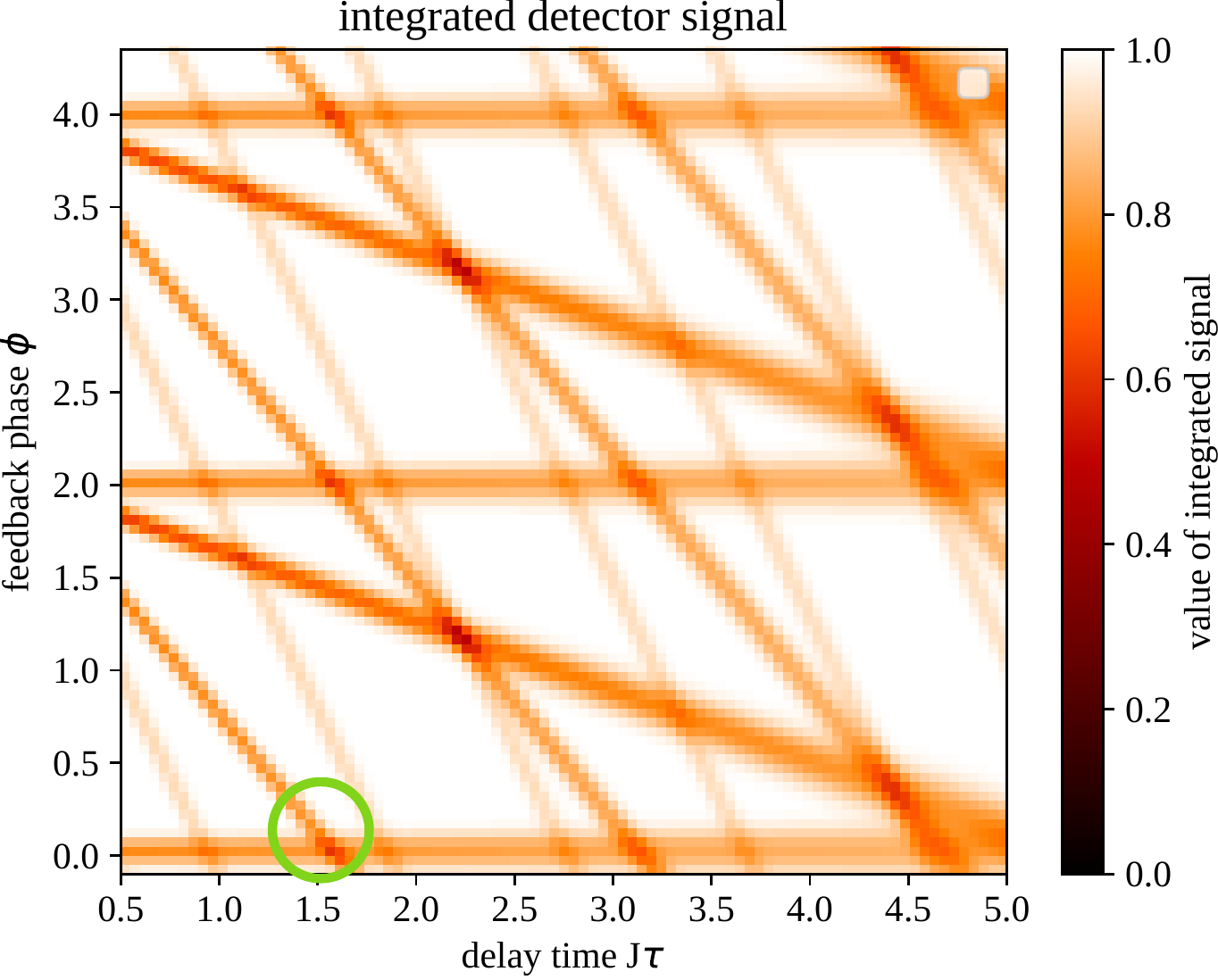}
		\caption{Stability landscape in the $\phi$-$\tau$ plane for an isotropic Heisenberg chain with nearest neighbor interaction and $N=4$ sites. The plot depicts the detector signal after a finite integration time $T$. Darker regions indicate a higher survival probability in the spin chain while brighter regions show that the excitation has been completely lost into the environment and detected. Broadening of the lines stems from finite calculation times, as mentioned in the main text. The periodic reappearance of the lines is due to the inherent $2\pi$-periodicity of the feedback phase $\phi$: Each stability condition is fulfilled once within every interval of $\phi \in [2\pi n, 2\pi (n+1))$, $n \in \mathbb{N}^+$. The green circle marks the intersection point of the parameter set $\phi_c$, $\tau_c$ in Fig.~\ref{fig3}. Parameters for this plot are $\Gamma=0.24$, $J=0.1$.}
		\label{fig4}
	\end{figure}
	%
	%
	\subsubsection{Stability planes and trapping conditions} \label{subsec:stability_planes}
	For the case of the many-body system under feedback, the conditions for the trapping to take place differ significantly from the case of a single two-level system. 
	We shortly repeat the distinguishing properties of a single two-level system which couples directly to the structured reservoir: Such a system never shows Rabi oscillations independently on the chosen phase and delay time, and population trapping only occurs at $\phi=\omega_0\tau=2\pi n$ with $n$ integer, i.e. in the interval $[0,2\pi)$ only one phase allows population trapping.

	This is significantly different in our system.
	To illustrate this, we plot in Fig.~\ref{fig4} the survival probability of the excitation in a chain of $N=4$ sites in the $\phi$-$\tau$-plane. 
	It depicts the time integrated detector signal, meaning that darker regions indicate a higher amount of trapped excitation within the system while brighter regions show that the excitation has been completely lost to the environment. Thus, all critical parameter sets $\phi_c$, $\tau_c$ for which trapping conditions exist are visible as lines in this plane.
	Note that the lines broaden out for two reasons: First, for the regions close to the critical parameters, $\phi \to \phi_c$, $\tau \to \tau_c$ and $\tau\Gamma \gg 1$, no trapping condition exists, however the feedback signal strongly slows down the dissipation into the environment. One could call these regions effectively-stable, which means the convergence time $T_c$ polynomially grows. As our numerical basis limits the total integration time, the stability lines broaden in Fig.~\ref{fig4} due to finite calculation times only. Also, note that for a fixed integration time $T$, the areas around the $\phi_c$-lines additionally broaden out with increasing delay time due to the convergence time strongly increasing with increasing $\tau$.

	Despite these obvious numerical limitations, we find in the interval $[0,2\pi)$ several conditions for $\phi_c$ which lead to population trapping, and the number of possible $\phi_c$ depends in strong contrast to the single two-level emitter case on $\tau$. 
	The reason for this is the interaction dynamics within the chain which imposes new conditions for the critical feedback phase $\phi_c$. 
	Additionally, in Fig.~\ref{fig4}, the dependency of the survival probability on $\tau$ for a fixed coupling strength $\Gamma$ becomes visible. The population trapping clearly decreases with an increasing delay time. This observation agrees with the behavior of the single two-level system with feedback and is due to the fact that the system loses excitation both to the feedback loop and to the waveguide constantly. If the signal travels very long through the feedback loop, only a small amount of excitation is left in the chain and the feedback-induced quantum interference between feedback-loop gain and waveguide-loss can only trap a small amount of excitation in the chain. This observation also explains that the higher the decay rate $\Gamma$, the smaller the survival probability for a fixed $\tau$.
	\begin{figure}[b!] 
		\centering
		\includegraphics[width=0.38\textwidth]{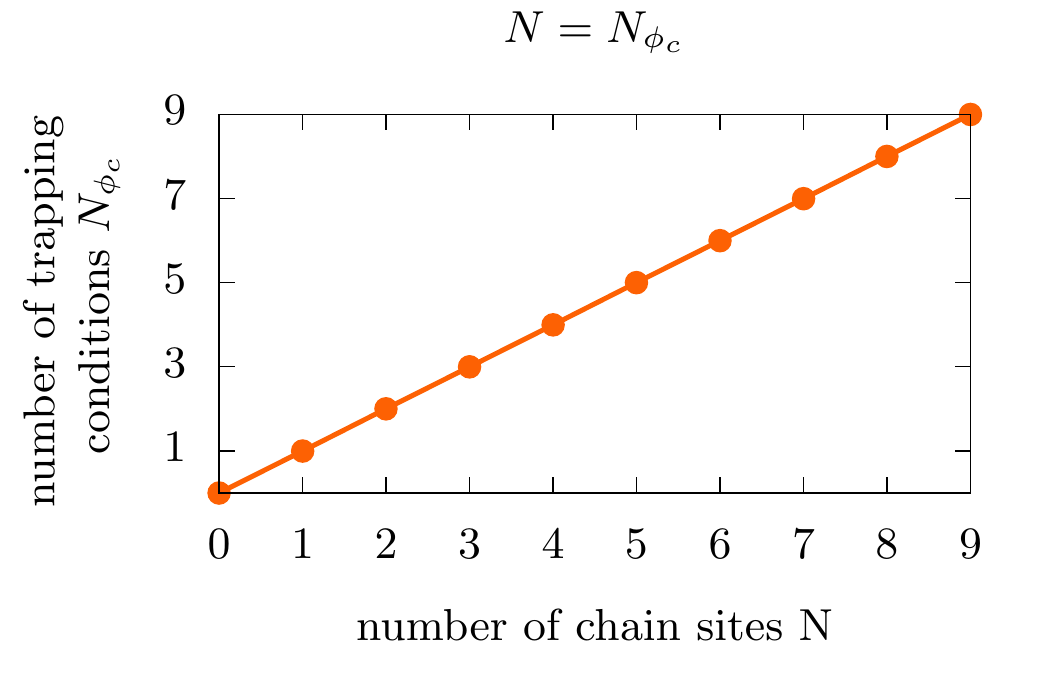}
		\caption{Plot of the maximum number of possible trapping conditions for the feedback phase $N_{\phi_c}$ within one interval $\phi \in [0,2\pi)$ as a function of the number of sites $N$ in the chain. Strikingly, it holds that $N_{\phi_c}$ = $N$. Scanning the possible population configurations allows to access the participating number of sites within the chain.}
		\label{fig5}
	\end{figure}
	The many-body system inherits nevertheless the $\phi=2n\pi$ stability from the single two-level case, which is visible as a horizontal line in Fig.~\ref{fig4}. Thus, here it holds that $\phi_c(\tau)=$ const., $\phi_c$ does not depend on $\tau$. Note that we assume a site independent system frequency $\omega_0$.
	For other phase choices, in the case of a many-body system under feedback, additional lines appear in the stability plane where it holds that $\phi_c=\phi_c(\tau)$. This dependency of the feedback phase on the delay time is an entirely new phenomenon compared to the well-investigated case of the single two-level system. 
	\begin{figure}[t!] 
		\centering
		\includegraphics[width=0.49\textwidth]{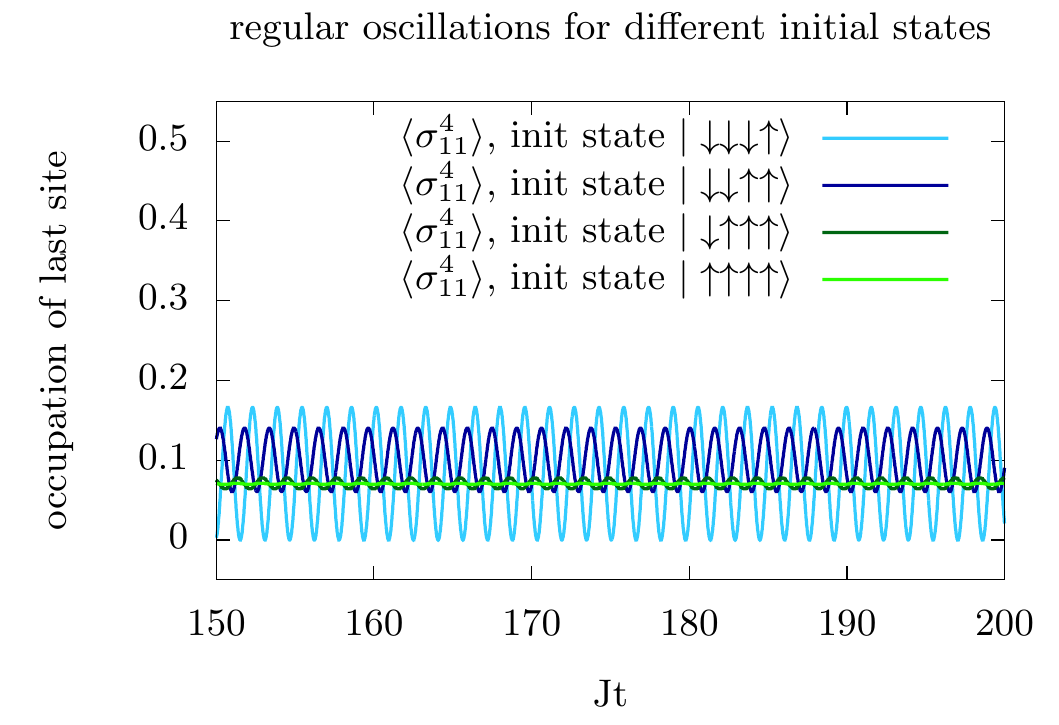}
		\caption{Regular oscillations for different initial states and initial numbers of excitations in a chain of $N=4$ sites. The amplitude decreases with an increasing number of excited states. The oscillations, however, remain regular and periodic. Parameters for this plot are $\Gamma = 0.24$, $J=0.1$.}
		\label{fig6}
	\end{figure}
	Due to the inherent periodicity of the phase, each of these additional lines appears once within every interval of $\phi \in [2\pi n, 2\pi (n+1))$, $n \in \mathbb{N}^+$, which means that the lines reappear periodically in the stability plane.
	We explain this $\tau$-dependency of the trapping conditions with the fact that the quantum  many-body system allows intrinsically for more coherent excitation exchange, and as the excitation is swapped back and forth in between the sites a phase is picked up which is intrinsically dependent on $J$ but does not change when we rescale the time. 

	This mechanism allows to extract via the integrated detection signal an estimate of the chain length of the participating sites. The number of possible population trapping conditions $N_{\phi_c}$ grows linearly with the number of sites, in fact, outside the points of degeneracy, the number of trapping condition equals the number of sites in the chain, $N_{\phi_c} = N$.
	This is a remarkable result of our study, as the detection signal reveals indirectly a 
	decisive quantum spin property unambiguously.
	If a point of degeneracy is chosen, furthermore, we find the highly non-trivial steady
	state of stabilized Rabi oscillations within the chain without any dephasing and dissipation although we simulate an open quantum system.
	This is discussed in the next section.
	%

	\subsubsection{Robustness of stabilized Rabi oscillations}
	%
	Investigating the steady-state behavior for different feedback phases and time delays, we observe three possibilities: (i) in the long-time limit all excitation of the chain is lost, (ii) all single site occupation densities in the chain are finite and constant, and (iii) the total excitation in the chain remains constant and finite but the densities oscillate. 
	Case (i) is the rule, not the exception, as most delay times and phases do not allow a non-trivial steady-state in combination with the quantum spin chain dynamics but will lead to a complete lost of excitation to the environment.
	Case (ii) is found where a feedback phase and delay time allow for population trapping, and a finite amount of excitation is found in the non-equilibrium steady-state.
	If however, degeneracy points are chosen for which the system provides two or more population trapping phases, a highly non-trivial steady-state is the result, namely (iii). 
	At these intersection of the stability lines, or degeneracy points, stabilized oscillations within the chain occur and a periodic, time-dependent steady-state is created.
	These steady states differ however in coherence and relative phase shifts between the trapped occupation densities $\langle \sigma_{11}^n \rangle_{\textup{tr}}$ at different intersection points. 
	An example of a very regular, time-reversible oscillation pattern is displayed in Fig.~\ref{fig3} and appears at a certain intersection which is marked in Fig.~\ref{fig4} with a green circle. 
	Characteristic for this non-equilibrium steady-state is the conservation of the excitation, thus:
	\begin{align}
	\sum_{n=1}^N \langle \sigma_{11}^n(t) \rangle_{\textup{tr}} \stackrel{!}{=} \textup{const}.
	\end{align}
	The same condition holds obviously for a closed chain. The main result of our study is the induced, synchronized and constant excitation within the chain although the system is open.
	This holds for different decay strengths $\Gamma$ and feedback delay times $\tau$, as well as feedback phases $\phi$, and is a generic feature of such a system.
	Here, the enabling factor is destructive interference at the entry point between the outgoing emission into the waveguide and the incoming feedback-signal. 
	Both re-excitation and de-excitation take place while applying the time-evolution operator of Eq.~\eqref{eq:u}. If the trapped occupation probabilities remain constant, as it is the case in two-level physics, this application will leave the matrix-product state unchanged.
	Therefore applying the MPO does not change the MPS although the spin chain couples dissipatively to a vacuum bin and a feedback bin.
	For the many-body system, this is the case if all occupation densities remain constant (case (ii)). 
	Contrary to the two-level physics, in case of the trapped Rabi-oscillations (case (iii)), we observe periodic changes when applying the MPO on the MPS. 
	This Floquet driving is a remarkable property of the many-body system and leads to the aforementioned regular oscillations without any decaying behavior. 
	In this section, we discuss additionally the robustness of this features.

	In Fig.~\ref{fig6}, the population trapping-induced oscillation within the chain is depicted for different initial states and number of excitations in the chain. 
	We clearly see that the effect is not limited to the single-excitation regime. 
	In contrast, the oscillating, time-periodic steady-state exists for different excitations and
	is a quite generic feature of the feedback-driven quantum spin chain. 
	However, the amplitude of the oscillations is reduced for larger numbers of excitations.
	This is displayed in Fig.~\ref{fig6}: The amplitude of the Rabi oscillations reaches its maximum for a single initial excitation (light blue line) and strongly decreases with an increasing number of initial excitation (e.g. quadruply-excited initial state, light green line). 
	We remark that this behavior is qualitatively independent of the location of the initial excitation within the chain.
	The explanation for the dependence of the amplitude on the initial number of excitations lies in the dynamics of the chain up to the first interaction with its own feedback signal. 
	The higher oscillations occurring in this first time interval $t\in[0,\tau]$, the higher the amplitude of the stabilized Rabi oscillations is in the long run. If the chain is initialized with a single excitation - no matter at which site in the chain - the oscillation of the occupation densities during this initial time interval has the highest amplitude, since the inversion of the individual site dynamics is not blocked by additional excitations. This amplitude decreases with an increasing number of excited sites and the oscillations in this first time interval $t\in[0,\tau]$ become increasingly irregular.
	Also, Fig.~\ref{fig4} shows that the total amount of trapped excitation in the chain is maximal at the intersection points, thus at the points where Rabi-oscillations occur.
	\section{Conclusion}\label{sec:conclusion}
	Contrary to the dominant Markovian approach for open spin chains, we investigate a Heisenberg spin chain with nearest neighbor interaction embedded into a non-Markovian structured reservoir. 
	This consists of a semi-infinite waveguide which feeds the emitted signal back into the chain.
	Thus, we extend the application of quantum feedback control, which is well investigated for few-level systems, on a many-body system. 
	We show that due to the many-body interactions, new trapping conditions arise with the feedback phase $\phi$ depending on the chosen delay time $\tau$. 
	Due to the periodicity of the phase, the set of trapping parameters are periodic in $[0,2\pi)$.
	Despite the complex interactions in the chain, the number $N_{\phi_c}$ of critical parameter sets $\phi_c$, $\tau_c$ for which trapping occurs within one interval is for most choices of $\tau$ equal to the number of sites $N$ in the chain. Also, we show that each specific parameter set $\phi_c$, $\tau_c$ relates to a specific state of the chain. We characterize these states with the numerical results for the occupation densities and show that at points in the $\phi$-$\tau$ plane where two trapping conditions hold, stable Rabi oscillations occur. Their amplitude is maximal for one single initial excitation in the chain. The total amount of trapped excitation $N_{tr}$ in the Rabi oscillations is maximal compared to all other trapped states. Our findings show that coherent-feedback is a promising way to study spin chains and other many-body quantum systems.
	
	\section*{Acknowledgements}
	We gratefully acknowledge the support of the Deutsche Forschungsgemeinschaft (DFG) through project B1 of the SFB 910.
	
	\begin{appendix}
		\section{Rotating frame transformation}
		\label{app:rotatingframe}
		We start with the the Hamiltonian in Eq.~\eqref{equ:model}, which reads (with $\hbar \equiv 1$):
		\begin{align}
		H = 
		&\sum_{i=1}^N \omega_0 \sigma_i^+\sigma_i^- + \int d\omega \omega b^\dagger (\omega) b (\omega)\nonumber\\
		&+\sum_{i=1}^{N-1} J\Big(\sigma^x_i\sigma^x_{i+1}+\sigma^y_i\sigma^y_{i+1}
		+\sigma^z_i\sigma^z_{i+1}\Big)\nonumber\\
		&+\int d\omega \left( G_{fb} (\omega) b^\dagger_N(\omega) \sigma_N^- + \textup{h.c.} \right)
		\end{align}
		In order to achieve a facilitating description for the numerical simulation, we transform this Hamiltonian into the rotating frame defined by its freely evolving part.
		For this, we use the unitary transformation with 
		\begin{equation}
		H'=U_1HU_1^\dagger -iU_1\partial_tU_1^\dagger
		\end{equation}
		where the unitary operator $U_1$ is defined as:
		\begin{equation}
		U_1=\exp{\left[ it \Big(\sum_{i=0}^N \omega_0 \sigma^+_i\sigma^-_i + \int d\omega \omega b^\dagger (\omega) b(\omega)\Big) \right]}
		\end{equation}
		This yields the transformed Hamiltonian $H'(t)$:
		\begin{align}
		H'(t) = 
		&\sum_{i=1}^{N-1}J\Big(\sigma^x_i\sigma^x_{i+1}+\sigma^y_i\sigma^y_{i+1}
		+\sigma^z_i\sigma^z_{i+1}\Big)\nonumber\\
		&+ \int d\omega \left( G_{fb} (\omega)\sigma^+_N b_N(\omega) e^{-i(\omega-\omega_s)t} + \textup{h.c.}\right)
		\end{align}
		Next, we again apply a unitary transformation, in order to shift the dependency of the delay time $\tau$ into the operators. This unitary operator $U_2$ is defined as:
		\begin{equation}
		U_2=\exp{\left[ -i\frac{\tau}{2}\int d\omega \omega b^\dagger(\omega) b(\omega) \right]}
		\end{equation}
		This yields:
		\begin{align}
		H'(t) = 
		&\sum_{i=1}^{N-1}J\Big(\sigma^x_i\sigma^x_{i+1}+\sigma^y_i\sigma^y_{i+1}
		+\sigma^z_i\sigma^z_{i+1}\Big)\nonumber\\
		&+ ig_0 \int d\omega \left( \sigma^+_N \big( b_N(\omega) e^{-i(\omega-\omega_s)t}\right.\nonumber\\
		& ~~~~~~~~~~~~~~~~~\left.- b_N(\omega) e^{-i(\omega-\omega_s)t} e^{i\omega \tau}\big) + \textup{h.c.}\right)
		\end{align}
		We define time dependent reservoir operators $b^{(\dagger)}(t)$ with
		\begin{equation}
		b(t) = \frac{1}{\sqrt{2\pi}}\int d \omega b(\omega) e^{-i(\omega-\omega_s)t}
		\end{equation}
		for which the following commutation relations hold:
		\begin{equation}
		[b(t),b^{\dagger}(t')]= \delta(t-t')
		\end{equation}
		With this, we arrive at the transformed Hamiltonian $H''$:
		\begin{align}
		H''(t) = 
		&\sum_{i=1}^{N-1}J\Big(\sigma^x_i\sigma^x_{i+1}+\sigma^y_i\sigma^y_{i+1}
		+\sigma^z_i\sigma^z_{i+1}\Big)\nonumber\\
		&+ i \sqrt{\Gamma}\Big(b_N(t)- b_N(t-\tau)e^{i\phi} \Big) \sigma_N^+ \nonumber\\
		&- i \sqrt{\Gamma}\Big(b_N^\dagger(t)- b_N^\dagger(t-\tau)e^{-i\phi} \Big) \sigma_N^-
		\label{equ:transformedhamilton}
		\end{align}
		with the feedback phase $\phi = \omega_s\tau$.
		
		\section{Quantum stochastical Schr\"odinger equation (QSSE)}
		\label{app:qsse}
		We use the picture of the quantum stochastical Schr\"odinger equation as basis for our numerical systems. Thus, we introduce time discrete quantum noise operators which include the interaction with the reservoir at one time step with a stochastical, continuous description \cite{Pichler2016,Nemet2019}
		\begin{align}
		\Delta B^{(\dagger)} (t_{k}) = \int_{t_k}^{t_{k+1}} dt'b^{(\dagger)}(t')
		\label{equ:qsseoperators}
		\end{align}
		with the following commutation relations:
		\begin{align}
		[B (t_{k}),&B^{\dagger} (t_{j})]= \nonumber\\
		&=\int_{t_k}^{t_{k+1}}dt \int_{t_j}^{t_{j+1}} dt' \delta(t-t') \nonumber\\
		&=\Delta t\delta_{kj}.
		\end{align}
		Note that $B^{(\dagger)} (t_k)$ and $B^{(\dagger)} (t_{k-l})$ only commute for $\Delta t = t_{k+1}-t_k < \tau$.\\
		The time evolution operator is defined as:
		\begin{equation}
		U(t)=\hat{T}\exp\left(-i\int_{t_0}^{t}H'' (t') dt'\right).
		\end{equation}
		We introduce the basis states \cite{Pichler2016}
		\begin{equation}
		\ket{i_p}=\frac{(\Delta B^\dagger (t_k))^{i_p}}{\sqrt{i_p!\Delta t^{i_p}}} \ket{\textup{vac}},
		\label{equ:basis}
		\end{equation}
		where $i_p$, $p$ integer, denotes the number of excitations present in the Fock state of the $k$th time interval $\ket{i_p}$.\\ 
		Writing Eq.~\eqref{equ:transformedhamilton} in the basis of the noise operators enables us to define a discretized time evolution operator $U(\Delta t)$ where we may drop the time evolution operator $\hat{T}$ for equidistant time steps $\Delta t=t_{k+1}-t_k$:
		\begin{align}
		U&(t_{k+1},t_k)= \nonumber\\
		&=\exp\Bigg[\sum_{i=1}^{N-1}J\Big(\sigma^x_i\sigma^x_{i+1}+\sigma^y_i\sigma^y_{i+1}
		+\sigma^z_i\sigma^z_{i+1}\Big)\nonumber\\
		&+\sqrt{\Gamma} \Big(\Delta B_N(t_{k})-\Delta B_N(t_{k-l}) e^{i\phi} \Big)\sigma_N^+ \nonumber\\
		&-\sqrt{\Gamma} \left(\Delta B^\dagger_N(t_{k}) - \Delta B^\dagger_N(t_{k-l}) e^{-i\phi} \right)\sigma_N^-\Bigg]
		\end{align}
		for $k\in[0,N_T-1]$ as integer of the time steps.
		Here, $t_k$ denotes the $k$th time step, while $t_{k-l}=(k-l)\Delta t$ denotes the time delayed by $\tau$, thus $t_k-\tau$ and $\phi=\omega_0\tau$ denotes the feedback phase.
		With this, we are able to use the QSSE operators defined in Eq.~\eqref{equ:basis} as the basis for the numerical non-Markovian time evolution. 
		
		\section{Matrix Product States}
		\label{app:mps}
		In order to compute the time evolution, we make use of tensor network methods by describing the state of the system and of the reservoir numerically as a matrix product state (MPS). Using the QSSE operators defined in Eq.~\eqref{equ:basis} as the numerical basis means that the corresponding Hilbert space scales with the integration time and thus becomes very large. Here, a time evolution based on the well established tensor network method MPS called tMPS  \cite{Schollwock2005,Schollwock2011,Vidal2003,Paeckel2019,Orus2008,White2004,White1993,Verstraete2004} allows for an efficient truncation of the Hilbert space.\\
		Central to this method is the expansion of the state vector coefficient into a matrix product state. For low dimensional few-level systems, the state of the system and the reservoir may be written into one single MPS - however, in case of a many-body system, this algorithm gets too demanding. Here, our method is the usage of a two dimensional MPS: In addition to the non-Markovian reservoir, our model also involves the spin chain as a quantum many body system with spacial interaction. Using the singular value decomposition, we expand the state vector coefficients both of system and reservoir into separated matrix product states \cite{Schollwock2011, Vidal2003, Paeckel2019}.
		The total wave vector reads as: 
		\begin{align}
		\ket{\psi(t_k)}=\nonumber &\sum_{\stackrel{n_1 \dots n_N}{= 0,1}} c_{n_1 \dots n_N} \ket{n_1\dots n_N} \\
		&\otimes 
		\sum_{k_1 \dots k_{N_T}} c_{k_1 \dots k_{N_T}} \ket{k_1\dots k_{N_T}}
		\end{align}
		with the expanded coefficients written with tensors $A$:
		\begin{align}
		c_{n_1 \dots n_N}&=A_{n_1}\cdot A_{n_2} \dots A_{n_N}\label{eq:mps1}\\
		c_{k_1 \dots k_{N_T}}&=A_{k_1}\cdot A_{k_2} \dots A_{k_{N_T}}\label{eq:mps2}
		\end{align}
		where the index $n_i$ is the physical index of the $i$th site in the chain and $k_j$ the index of the state of the reservoir at the $j$th time step. Thus, Eq.~\eqref{eq:mps1} describes the wave vector of the many-body system as MPS, while Eq.~\eqref{eq:mps2} the one of the reservoir.\\
		These two MPS contain the physical information of the system as well as of the state of the reservoir at every time step. They consist of $N_T$ respectively $N$ connected tensors called bins, where $N_k=\frac{T}{\Delta t}$ is the total number of time steps and $N$ the number of sites in the chain. Thus, in the reservoir MPS, every bin represents the state of the reservoir at one time step, while in the spin chain MPS, each bin represents one site. The two MPS are stuck together at the $N$th chain bin and the $k$th time bin, where the interaction between the many-body system and the reservoir occurs. \\
		Using this form allows not only for the preservation of the state of the reservoir at every time step, but more importantly for the efficient truncation of the Hilbert space: The singular values of the decomposed wave vector matrices represent the entanglement in between the many-body system, between reservoir and spin-chain as well as in between the state of the reservoir at different time steps. Truncating their entries during the decomposition process, thus setting them to zero below a given cutoff-threshold, reduces the computed part of the Hilbert space efficiently while loosing only the paths with negligible probabilities.\\
		
		\section{Employing tMPS for coherent self-feedback}
		\label{app:algorithm}
		In order to compute the $k$th time step, we contract the $N$th chain bin, the $k$th time bin initialized in a vacuum state and the $t_{k-l}$th bin containing the feedback signal. The time evolution operator $U(t_{k+1},t_k)$ is expanded into a matrix product operator (MPO), and the time evolution of one time step is computed as $\ket{\psi(t_{k+1})}=U(t_{k+1},t_k)\ket{\psi(t_k)}$, which means the MPO is multiplied into the MPS of the spin chain where the last site contains all relevant information for the interaction with the reservoir at the present time step. \\
		After applying the MPO, we decompose the tensor again, shift the bins back to their original position in the chain, move and contract the bins of the $(k+1)$th time step and so forth. Care has to be taken to keep the orthogonality center at the right position in order to preserve the entanglement information correctly.\\
		This algorithm enables us to efficiently simulate a quantum many body system under the influence of coherent self-feedback.\\
	\end{appendix}
	
	\bibliography{MyCollection}
	
\end{document}